\documentclass[sigconf,review=false]{acmart}

\usepackage{xspace}
\newcommand\cop{{\sc Copy}\xspace}
\newcommand\del{{\sc Delete}\xspace}

\newcommand\rep{{\sc Replace}\xspace}

\newcommand\copline{{\sc CopyLine}\xspace}
\newcommand\delline{{\sc DeleteLine}\xspace}

\newcommand\repline{{\sc ReplaceLine}\xspace}
\newcommand\swaline{{\sc SwapLine}\xspace}

\newcommand\copstate{{\sc CopyStatement}\xspace}
\newcommand\delstate{{\sc DeleteStatement}\xspace}

\newcommand\repstate{{\sc ReplaceStatement}\xspace}
\newcommand\swastate{{\sc SwapStatement}\xspace}

\usepackage{xfrac}

\usepackage{listings}
\usepackage{color}

\usepackage{stfloats}
\usepackage{booktabs}
\usepackage{multirow}

\AtBeginDocument{
  \providecommand\BibTeX{{
    \normalfont B\kern-0.5em{\scshape i\kern-0.25em b}\kern-0.8em\TeX}}}
    
\setcopyright{none}
\renewcommand\footnotetextcopyrightpermission[1]{}

\begin{document}

\title{On the Utility of Marrying GIN and PMD for Improving Stack Overflow Code Snippets}

\author{Sherlock A. Licorish}
\affiliation{
\institution{Department of Information Sciences, University of Otago}
\city{Dunedin}
\country{New Zealand}}
\email{sherlock.licorish@otago.ac.nz}

\author{Markus Wagner}
\affiliation{
\institution{School of Computer Science, The University of Adelaide}
\city{Adelaide}
\country{Australia}}
\email{markus.wagner@adelaide.edu.au}

\renewcommand{\shortauthors}{Licorish and Wagner}

\begin{abstract}

Software developers are increasingly dependent on question and answer portals and blogs for coding solutions. While such interfaces provide useful information, there are concerns that code hosted here is often incorrect, insecure or incomplete. Previous work indeed detected a range of faults in code provided on Stack Overflow through the use of static analysis. Static analysis may go a far way towards quickly establishing the health of software code available online. In addition, mechanisms that enable rapid automated program improvement may then enhance such code. Accordingly, we present this proof of concept. We use the PMD static analysis tool to detect performance faults for a sample of Stack Overflow Java code snippets, before performing mutations on these snippets using GIN. We then re-analyse the performance faults in these snippets after the GIN mutations. GIN's RandomSampler was used to perform 17,986 unique line and statement patches on 3,034 snippets where PMD violations were removed from 770 patched versions. Our outcomes indicate that static analysis techniques may be combined with automated program improvement methods to enhance publicly available code with very little resource requirements. We discuss our planned research agenda in this regard.

\end{abstract}

\vspace{-2mm}\keywords{Static analysis, Genetic improvement, Hybridisation}
\maketitle
\pagestyle{plain}
\sloppy 

\vspace{-2mm}\section{Introduction/Motivation}
Websites hosting code online such as Stack Overflow and HackerRank have become the cornerstone for software developers seeking solutions to their coding challenges ~\cite{meldrum0}. These portals are particularly useful as they allow the community to openly critique solutions. While this mechanism is anticipated to help with improving code quality, evidence has shown that many faults remain in code available in online portals~\cite{meldrum}. Stack Overflow code, in particular, is extensively reused, at times introducing unsuspecting vulnerabilities in the systems where such code is copied~\cite{lotter}.

In helping to remedy faulty code, static analysis is used extensively for understanding the quality of code on Stack Overflow and other online portals. For instance, PMD and SpotBugs have provided insights around how much contributors adhere to code readability, reliability, performance and security rules~\cite{meldrum,gkortzis}. These tools may help the software engineering community to quickly understand the quality of code, and whether or not provided solutions conform to coding standards. However, they are ignored at times due to the large amount of rule violations that are returned. In fact, such tools may be combined with automated program improvement techniques for improving code quality, where faults detected through the use of static analysis may be remedied with automated patches, thereby aiding the community. That said, while exciting, automated program improvement as a discipline faces its own challenges in terms of the intelligent navigation of a search space of probable changes and synthesizing appropriate code for patching~\cite{harrand2019neutral}.

Notwithstanding the challenges apparent in static analysis and automated program improvement techniques, we believe that they may be combined to good effect for helping to improve code available online. In demonstrating this proof of concept, we use the PMD static analysis tool ~\cite{pmd} to detect performance faults for a sample of Stack Overflow Java code snippets, before performing mutations on these snippets using GIN~\cite{gin}. We then re-analyse the performance faults in these snippets after the GIN mutations, observing indicators that static analysis techniques may be combined with automated program repair methods to good effect.

Our contributions in this poster paper are twofold. We demonstrate the utility of static analysis and automated program improvement, and more granular, we provide empirical evidence for how PMD and GIN may be combined for code improvement. We also propose several future research directions, in presenting an agenda for how the two domains considered may be further explored.

\section{Initial Study}

\subsection{Data}\label{sec:data}
We have used the 8010 snippets that were provided by~\cite{meldrum}. The data were extracted from Stack Overflow for 2014, 2015, and 2016 and were said to represent suitably long compilable code from answers where a high level of reuse was evident. Java code was studied due to its popularity, and various forms of preprocessing were done to ensure a reliable dataset. For instance, code snippets were checked for the `import', `package', or `class' keyword and saved unmodified. Code snippets that did not contain these words were encased in a public class structure, and files were saved with the .java extension with a unique name. 

As GIN's samplers target the modification of methods, we discard files that do not contain any methods. Moreover, we discard files that do not compile or that contain Java features that are currently not supported by GIN, e.g. certain multi-threading concepts. 
The resulting 3034
independent files contain 3607 methods. This number is worth highlighting, as most analyses of program improvement spaces to date consider only single programs.

To perform our static analyses, we employ PMD~\cite{pmd}. For Java, it has 324 rules {(\small\texttt{rulesets/internal/all-java.xml})} organised in eight sets: Best Practices, Code Style, Design, Documentation, Error Prone, Multi-Threading, Performance, and Security. Analysing the 3034 files resulted in 30,668 PMD violations.

Table~\ref{orig3034} shows a summary of the 30,668 PMD rule violations; 
in total, 135 different rules were violated. Table~\ref{orig3034top3} lists for each of the seven categories the three most frequently violated rules. We spell out the rules to provide the reader with an idea of the types of rules that PMD contains. PMD's documentation also contains longer descriptions as well as examples and suggestions for mitigation.

\renewcommand*{\arraystretch}{0.95}

\begin{table}[!t]
\caption{PMD output on the original 3034 code snippets: 30,668 rule violations are found by 135 (of 324) rules.}
\label{orig3034}\vspace{-3mm}
\begin{tabular}{lrr}
\toprule
PMD ruleset    & \shortstack[c]{total number\\of violations} & \shortstack[r]{different rules\\violated (total)} \\\midrule
Code Style    CS           & 16832                     & 31  (64)                               \\
Documentation   DOC          & 6292                      & 3  (5)                                \\
Best Practice   BP           & 3557                      & 23  (57)                               \\
Design          DES          & 2785                      & 26 (48)                           \\
Error Prone     EP           & 778                       & 31  (103)                               \\
Performance     PER          & 396                       & 17  (32)                               \\
Multi-Threading  MT           & 28                        & 4 (11)                          \\
Security SEC & 0 & 0 (4)       
\\\bottomrule
\end{tabular}\vspace{-1mm}
\end{table}

\begin{table*}[bp]
\vspace{0.5mm}
\caption{PMD output on the original 3034 code snippets: three most-frequently violated rules per ruleset.  Sorted by the number of violations.
Examples of rule instantiations are shown as \emph{<\hspace{0.2mm}...>}.
}
\label{orig3034top3}\vspace{-3mm}\small{\setlength{\tabcolsep}{1pt}
\begin{tabular}{llrl}
\toprule
rule & ruleset & count & description \\ \midrule
CommentRequired                 & DOC & 6131 & Class comments are required.                                                                                                                                                 \\
MethodArgumentCouldBeFinal      & CS  & 4077 & Parameter `givenString' is not assigned and could be declared final.                                                                                                         \\
NoPackage                       & CS  & 3034 & All classes, interfaces, enums and annotations must belong to a named   package.                                                                                             \\
ShortVariable                   & CS  & 2496 & Avoid variables with short names like \emph{<s>}.                                                                                                                                     \\
UseUtilityClass                 & DES & 1722 & All methods are static.  Consider   using a utility class instead. 
\\
SystemPrintln                   & BP  & 1718 & System.out.println is used.                                                                                                                                                  \\
LawOfDemeter                    & DES & 672  & Potential violation of Law of Demeter (method chain calls).                                                                                                                  \\
UseVarargs                      & BP  & 461  & Consider using varargs for methods or constructors which take an array   the last parameter.                                                                                \\
AvoidLiteralsInIfCondition      & EP  & 435  & Avoid using Literals in Conditional Statements.                                                                                                                              \\
AvoidReassigningParameters      & BP  & 254  & Avoid reassigning parameters such as \emph{<x>}.                                                                                                                                    \\
UseStringBufferForStringAppends & PER & 162  & Prefer StringBuilder (non-synchronized) or StringBuffer (synchronized)   over += for concatenating strings.\hspace{-2mm}           \\
BeanMembersShouldSerialize      & EP  & 93   & Found non-transient, non-static member. Please mark as transient or   provide accessors.                                                                                    \\
CommentSize                     & DOC & 92   & Comment is too large: Line too long.                                                                                                                                         \\
UncommentedEmptyMethodBody      & DOC & 69   & Document empty method body.                                                                                                                                                  \\
AddEmptyString                  & PER & 57   & Do not add empty strings.                                                                                                                                                    \\
CyclomaticComplexity            & DES & 52   & The method \emph{<translation(String)>} has a cyclomatic complexity of \emph{<13>}.                                                                                                         \\
UseLocaleWithCaseConversions    & EP  & 45   & When doing a String.toLowerCase()/toUpperCase() call, use a Locale.                                                                                                          \\
AppendCharacterWithChar         & PER & 39   & Avoid appending characters as strings in StringBuffer.append.                                                                                                               \\
AvoidSynchronizedAtMethodLevel  & MT  & 20   & Use block level rather than method level synchronization.                                                                                                                    \\
AvoidUsingVolatile              & MT  & 4    & Use of modifier volatile is not recommended.                                                                                                                                \\
UnsynchronizedStaticFormatter   & MT  & 3    & Static Formatter objects should be accessed in a synchronized manner.                                                                                     
\\\bottomrule                                           
\end{tabular}}
\end{table*}

\subsection{Sampling of the Edit Space}\label{sec:sampling}

To create large numbers of patched code, we use GIN~\cite{gin}, an extensible and modifiable toolbox for search-based experimentation with code. GIN automatically transforms, builds, and tests Java projects. 
In particular, we employ GIN's RandomSampler: it randomly generates a patch (which is composed of a given number of individual edits), it applies that patch, then tests the resulting source, and finally returns the result. RandomSampler does not perform a random walk or any iterated search via a sequence of patches, but it always takes the original file as the starting point for the application of the next patch. This enable the characterisation of neighbourhoods in the program space.

In this study, we sample small patches, i.e., patches that contain only one edit of the following eight: \delline{}, \repline{}, \copline{}, and \swaline{}; and \delstate{}, \repstate{}, \copstate{}, and \swastate{}. 
We do not explore sequential edits, other edit types, and of more complex program transformations, as these are beyond the scope of this short article.

We generate 10,000 patches with one line edit, and 10,000 patches with one statement edit. 
As the files are relatively small (14.7 lines on average), there is a chance of randomly sampling the very same patch again, such as the deletion of a particular statement. Therefore, instead of 20,000 unique patches, only 17,986 patches of the original code snippets were recorded. 

Among these, 5640 (31.4\%) are compilable, which is aligned with earlier observations made, e.g., by \citet{langdon2017fragile}, that code is not particularly fragile. Moreover, we observe that the likelihood for a statement-level edit to compile (45.1\%) is more than twice as high as it is for a line-level edit (19.8\%). 

\subsection{Static Analysis with a Performance Focus}
\label{saperf}

Earlier work singled out code performance issues as serious~\cite{meldrum}, thus, we now limit our proof of concept investigations to PMD's 32 performance rules ({\small\texttt{category/java/performance.xml}}). According to PMD, these are ``Rules that flag suboptimal code''.

Given the union of the original code snippets and the patched ones, PMD finds 3121 performance issues in 1203 of the original 3034 files. 
Given these 1203 files, we focus on 1185 of them (with a total of 1915 performance issues): these are 237 (with a total of 349 performance issues) of the original code snippets that originally had performance issues, plus their 1185-237=948 patched versions (with a total of 1915-349=1566 performance issues) for which PMD reports performance issues, too. This allows us to observe the effect of patches to code with performance issues. 

Note that there is an additional set of files that we briefly characterise first: the 770 patched versions of the 237 files that no longer have any issues associated with them. For these 770 files, 547 times 1 issue is removed, 172x 2, 26x 3, 21x 4, 2x 5, and 2x 8 issues. 
While this undifferentiated view at GIN's removal of performance issues seems fantastic at first, we need to note that {\footnotesize$\frac{712}{770}$}=92.5\% of the patched files do not compile, which is much above the expected average of 68.6\% (see in Section~\ref{sec:sampling}). This points us at a major issue: code that does not compile seems to pose a major challenge for PMD's performance-related rules, or for parts of its inner workings, such as its parser. 

Of the 58 edits that produce compilable code while removing all performance issues, 36 are \del{} edits that simply delete the offending code. Of the 22 other edits, most of them either replace or modify the offending code (see Listing~\ref{lst:patched} for an example), or they replace the loop with its body, which then no longer violates loop-based rules. Lastly, we have identified one case where PMD incorrectly does not advise of an actual rule violation.

\makeatletter
\let\orig@lstnumber=\thelstnumber
\newcommand\lstsetnumber[1]{\gdef\thelstnumber{#1}}
\newcommand\lstresetnumber{\global\let\thelstnumber=\orig@lstnumber}
\makeatother

\lstsetnumber{}
\begin{lstlisting}[float=tb,caption={Snippet C402521.java: \swastate{}(153,154) swaps the body of the loop with one of its statements, thus mitigating the AvoidArrayLoops violation. Shown is only the relevant part of the code.},label=lst:patched,mathescape=true]
// original $\lstresetnumber\setcounter{lstnumber}{23}$
for (int i = c+1; i < nums.length; i++) {
    b[j] = nums[i];
    j++;
}$\lstsetnumber{}$
// after applying the patch $\lstresetnumber\setcounter{lstnumber}{23}$
for (int i = c+1; i < nums.length; i++) 
    b[j] = nums[i];
\end{lstlisting}

Coming back to the files with performance issues, Table~\ref{orig237} shows the 349 issues that PMD has detected in the original 237 code snippets. From left to right, the table provides PMD's internal rule name, the count of how often it was triggered, and PMD's short description of the rule. Complementary to this, we show in Table~\ref{perfisssues1566} the distribution of performance issues in the derived patches that still exhibit performance issues. Among other, we can see that (i) the original files have {\footnotesize$\frac{349}{237}$}=1.47 performance issues on average, (ii) all patched versions together have {\footnotesize$\frac{739+827}{433+495+770}$}=0.92 performance issues on average, and (iii) the compilable, patched code has {\footnotesize$\frac{739}{453+58}$}=1.45 performance issues on average. 

In addition to this, we can observe that some edit types appear to attract or mitigate more rule violations than others. For example, the \cop{} edits attract disproportionally many violations, and the \del{} edits perform best against the AvoidInstantiatingObjectsInLoops violations. However, such observations need to be taken with a grain of salt, because (i) our sampling was not exhaustive but random, and (ii) we are lacking code tests. That said, in alignment with our focus, the marriage of PMD and GIN seems noteworthy.

\begin{table*}[]
\caption{PMD finds 349 performance-related rule violations in the 237 original code snippets.  
Examples of rule instantiations are shown as \emph{<\hspace{0.2mm}...>}.
}\vspace{-3mm}
\label{orig237}\small{\setlength{\tabcolsep}{1pt}
\begin{tabular}{lrl} \toprule
rule & count & description \\ \midrule
UseStringBufferForStringAppends     & 118 & Prefer StringBuilder (non-synchronized) or   StringBuffer (synchronized) over += for concatenating strings. \\
AddEmptyString                      & 54  & Do not add empty strings.                                                                                                                 \\
AppendCharacterWithChar             & 35  & Avoid appending characters as strings in StringBuffer.append.                                                                            \\
RedundantFieldInitializer           & 23  & Avoid using redundant field initializer for \emph{<i>}.                                                                       \\
AvoidInstantiatingObjectsInLoops    & 19  & Avoid instantiating new objects inside loops.                                                                                             \\
AvoidArrayLoops                     & 19  & System.arraycopy is more efficient.                                                                                                       \\
UseIndexOfChar                      & 12  & String.indexOf(char) is faster than String.indexOf(String).                                                                              \\
StringInstantiation                 & 11  & Avoid instantiating String objects; this is usually unnecessary.                                                                         \\
InefficientStringBuffering          & 9   & Avoid concatenating nonliterals in a StringBuffer/StringBuilder   constructor or append().                                               \\
AvoidUsingShortType                 & 8   & Do not use the short type.                                                                                                                \\
TooFewBranchesForASwitchStatement\hspace{-2.6mm}   & 7   & A switch with less than three branches is inefficient, use a if statement   instead.                                                     \\
IntegerInstantiation                & 6   & Avoid instantiating Integer objects. Call Integer.valueOf() instead.                                                                     \\
UselessStringValueOf                & 6   & No need to call String.valueOf to append to a string.                                                                                    \\
ConsecutiveAppendsShouldReuse       & 4   & StringBuffer (or StringBuilder).append is called consecutively without   reusing the target variable.                                    \\
InefficientEmptyStringCheck         & 4   & String.trim().length() == 0 / String.trim().isEmpty() is an inefficient   way to validate a blank String.                                \\
StringToString                      & 3   & Avoid calling toString() on String objects; this is unnecessary.                                                                         \\
InsufficientStringBufferDeclaration & 3   & StringBuilder has been initialized with size \emph{<16>}, but has at least \emph{<143>} characters appended.                                             \\
SimplifyStartsWith                  & 3   & This call to String.startsWith can be rewritten using String.charAt(0).                                                        \\
ConsecutiveLiteralAppends           & 2   & StringBuffer (or StringBuilder).append is called \emph{<3>} consecutive times with literals. \\
OptimizableToArrayCall              & 2   & This call to Collection.toArray() may be optimizable.                                                                                     \\
BooleanInstantiation                & 1   & Avoid instantiating Boolean objects; reference Boolean.TRUE/Boolean.FALSE or call Boolean.valueOf() instead.    
\\ \bottomrule
       
\end{tabular}}
\end{table*}

\begin{table*}[]\setlength{\tabcolsep}{4.6pt}\vspace{3mm}
\caption{PMD finds 1566 performance-related issues in the patched code: 739 in the 453 compilable snippets and 827 in the 495 snippets that do not compile. The leftmost column shows the statistics for the original 237 code snippets from Table~\ref{orig237} as a reference.}\vspace{-3mm}
\label{perfisssues1566}\small{
\begin{tabular}{r|l|rrrrrrrrr|rrrrrrrrr} \toprule
\multirow{3}{*}{\shortstack[c]{\vspace*{8mm}\ \\original\\code\\snippets}} & \multirow{3}{*}{\shortstack[c]{\vspace*{12mm}\ \\ rule}} & \multicolumn{9}{c|}{patched files that compile} & \multicolumn{9}{c}{patched files that do not compile}  \\ 
&                                     & total & \rotatebox{90}{\copline{}} & \rotatebox{90}{\delline{}} & \rotatebox{90}{\repline{}} & \rotatebox{90}{\swaline{}} & \rotatebox{90}{\copstate{}} & \rotatebox{90}{\delstate{}} & \rotatebox{90}{\repstate{}} & \rotatebox{90}{\swastate{}} & total & \rotatebox{90}{\copline{}} & \rotatebox{90}{\delline{}} & \rotatebox{90}{\repline{}} & \rotatebox{90}{\swaline{}} & \rotatebox{90}{\copstate{}} & \rotatebox{90}{\delstate{}} & \rotatebox{90}{\repstate{}} & \rotatebox{90}{\swastate{}} \\ \midrule
118        & UseStringBufferForStringAppends     & 247        & 30       & 23         & 17          & 30       & 63            & 20              & 35               & 29            & 262         & 32       & 14         & 7           & 12       & 114           & 18              & 50               & 15            \\
54         & AddEmptyString                      & 121        & 2        & 14         & 10          & 15       & 22            & 23              & 20               & 15            & 157         & 4        & 16         & 16          & 14       & 42            & 10              & 37               & 18            \\
35         & AppendCharacterWithChar             & 100        & 5        & 18         & 9           & 14       & 16            & 14              & 9                & 15            & 102         & 11       & 3          & 2           & 5        & 34            & 17              & 15               & 15            \\
23         & RedundantFieldInitializer           & 55         & 2        & 4          & 3           & 6        & 10            & 6               & 12               & 12            & 36          & 2        & 1          & 0           & 3        & 9             & 10              & 5                & 6             \\
19         & AvoidInstantiatingObjectsInLoops    & 26         & 2        & 3          & 3           & 4        & 4             & 1               & 7                & 2             & 74          & 5        & 3          & 3           & 10       & 18            & 3               & 16               & 16            \\
19         & AvoidArrayLoops                     & 27         & 3        & 2          & 2           & 2        & 7             & 7               & 1                & 3             & 45          & 2        & 5          & 0           & 2        & 16            & 5               & 9                & 6             \\
12         & UseIndexOfChar                      & 24         & 2        & 1          & 0           & 2        & 12            & 5               & 0                & 2             & 20          & 0        & 0          & 1           & 1        & 13            & 1               & 2                & 2             \\
11         & StringInstantiation                 & 28         & 3        & 2          & 2           & 5        & 5             & 7               & 2                & 2             & 19          & 0        & 1          & 0           & 0        & 8             & 6               & 3                & 1             \\
9          & InefficientStringBuffering          & 9          & 2        & 0          & 2           & 0        & 2             & 1               & 1                & 1             & 11          & 0        & 1          & 0           & 1        & 5             & 1               & 2                & 1             \\
8          & AvoidUsingShortType                 & 24         & 0        & 2          & 2           & 4        & 0             & 6               & 3                & 7             & 25          & 1        & 5          & 0           & 0        & 9             & 4               & 6                & 0             \\
7          & TooFewBranchesForASwitchStatement   & 12         & 0        & 0          & 0           & 1        & 3             & 4               & 2                & 2             & 18          & 0        & 2          & 0           & 0        & 3             & 4               & 4                & 5             \\
6          & IntegerInstantiation                & 8          & 0        & 0          & 2           & 1        & 1             & 1               & 2                & 1             & 14          & 1        & 0          & 0           & 1        & 7             & 3               & 2                & 0             \\
6          & UselessStringValueOf                & 14         & 0        & 1          & 0           & 1        & 8             & 1               & 1                & 2             & 7           & 0        & 3          & 0           & 0        & 4             & 0               & 0                & 0             \\
4          & ConsecutiveAppendsShouldReuse       & 12         & 1        & 2          & 2           & 0        & 5             & 2               & 0                & 0             & 18          & 5        & 2          & 0           & 0        & 5             & 1               & 4                & 1             \\
4          & InefficientEmptyStringCheck         & 6          & 0        & 1          & 0           & 2        & 0             & 0               & 0                & 3             & 3           & 0        & 0          & 0           & 0        & 2             & 1               & 0                & 0             \\
3          & StringToString                      & 3          & 0        & 1          & 1           & 0        & 0             & 0               & 1                & 0             & 2           & 0        & 0          & 0           & 0        & 2             & 0               & 0                & 0             \\
3          & InsufficientStringBufferDeclaration & 7          & 0        & 2          & 1           & 0        & 0             & 1               & 2                & 1             & 5           & 0        & 1          & 0           & 0        & 4             & 0               & 0                & 0             \\
3          & SimplifyStartsWith                  & 8          & 0        & 0          & 2           & 3        & 0             & 0               & 2                & 1             & 3           & 0        & 0          & 2           & 0        & 0             & 0               & 1                & 0             \\
2          & ConsecutiveLiteralAppends           & 3          & 0        & 0          & 0           & 0        & 2             & 0               & 1                & 0             & 2           & 0        & 0          & 0           & 0        & 0             & 1               & 0                & 1             \\
2          & OptimizableToArrayCall              & 4          & 0        & 0          & 0           & 0        & 2             & 2               & 0                & 0             & 2           & 0        & 0          & 0           & 0        & 2             & 0               & 0                & 0             \\
1          & BooleanInstantiation                & 1          & 0        & 0          & 0           & 0        & 0             & 0               & 0                & 1             & 2           & 1        & 0          & 0           & 0        & 1             & 0               & 0                & 0             \\ \midrule
349        & \textit{total rule violations}                               & 739        & 52       & 76         & 58          & 90       & 162           & 101             & 101              & 99            & 827         & 64       & 57         & 31          & 49       & 298           & 85              & 156              & 87    \\ \midrule 
\multicolumn{1}{r}{} & \multicolumn{1}{r|}{\textit{number of patched files:}}     & 453        & 28       & 42         & 36          & 54       & 82            & 67              & 71               & 73            & 495         & 47       & 39         & 16          & 37       & 161           & 60              & 82               & 53            \\ \bottomrule
\end{tabular}}
\end{table*}

\section{Conclusions and Future Directions }

Our outcomes show promise, and established the proof of concept: static analysis techniques and automated program improvement methods can be combined to enhance publicly available code. That said, there are several open areas that require further investigation.  

\vspace{-2mm}\paragraph{\textbf{Better Static Analysis.}}

\emph{(1) Mitigating False Positives and Trivial Warnings:} The sheer dimensionality of PMD's output requires mechanisms to establish potential false positives and eliminate trivial warnings. PMD's readability and reliability warnings have been show to be accurate~\cite{meldrum}, and should make a good starting point. However, SpotBugs may offer an alternative body of checks. \emph{(2) Improve Parsing:} PMD's effectiveness appears to be severely affected when code does not compile, hence possibly requiring a more robust approach to parsing. \emph{(3) Crowd-Sourcing Rules:} It may be possible to crowd-source further performance-related PMD rules by mining repositories and question-answering sites like Stack~Overflow; for example, \citet{baltes2020sodata} provide a potentially useful dataset.

\vspace{-2mm}\paragraph{\textbf{Better Automated Program Improvement.}}

\emph{(1) Biased Sampling:} As different edits result in different distributions of triggered rules, we conjecture that machine learning models (that take PMD output as input) can be used to bias the patch generation towards desired code properties. \emph{(2) Better Code Transformations:} While the traditional operators for search-based program modification (e.g., copy, delete, replace, and swap) seem inadequate at first to address the PMD rules violations that we have encountered here (see Section~\ref{saperf}), we can imagine scenarios where a single \rep{} or \del{} can resolve a violation, e.g., in certain cases of AddEmptyString and UseIndexOfChar. For other violations, e.g., OptimizabletoArrayCall and AvoidInstantiatingObjectInLoops, however, custom transformations appear to be necessary -- possibly, insights from the well-established field of code refactoring~\cite{fowler2018refactoring} can be beneficial. \emph{(3) Further Code Properties:} While our study here focuses almost exclusively on performance-related improvements, it is straightforward to change the focus to other sets of PMD rules, to other non-functional properties, and even to functional properties.

\bibliographystyle{ACM-Reference-Format}
\bibliography{sample-base}

\end{document}